\documentclass[aps,prd,twocolumn,superscriptaddress,amssymb,eqsecnum,showpacs,showkeyes,secnumarabic,graphics,floatfix,nofootinbib,tightenlines,longbibliography]{revtex4-1}

\usepackage{booktabs} % For better looking tables
\usepackage[margin=1in]{geometry} % To adjust margins
\usepackage{lipsum} % For dummy text

\usepackage{graphicx}
\usepackage{bm}
\usepackage{cancel}
\usepackage{epsfig}
\usepackage{dcolumn}
\usepackage{amsmath}
\usepackage{hhline}
\usepackage{enumerate}
\usepackage[utf8]{inputenc}
\usepackage[dvipsnames]{xcolor}
\usepackage[breaklinks=true,colorlinks=true,linkcolor=blue,urlcolor=Blue,citecolor=MidnightBlue,% PDF VIEW
bookmarks=true,bookmarksopenlevel=2]{hyperref}
\usepackage{bm}
\usepackage{amsmath}
\usepackage{amssymb}
\usepackage{subcaption}
\usepackage{academicons}

\begin{document}
\newcommand{\newc}{\newcommand}

\newc{\be}{\begin{equation}}
\newc{\ee}{\end{equation}}
\newc{\ba}{\begin{eqnarray}}
\newc{\ea}{\end{eqnarray}}
\newc{\bea}{\begin{eqnarray*}}
\newc{\eea}{\end{eqnarray*}}
\newc{\D}{\partial}
\newc{\ie}{{\it i.e.} }
\newc{\eg}{{\it e.g.} }
\newc{\etc}{{\it etc.} }
\newc{\etal}{{\it et al.}}
\newc{\lcdm}{$\Lambda$CDM }
\newc{\lcdmnospace}{$\Lambda$CDM}
\newc{\wcdm}{$w$CDM }
\newc{\plcdm}{Planck18/$\Lambda$CDM }
\newc{\plcdmnospace}{Planck18/$\Lambda$CDM}
\newc{\omom}{$\Omega_{0m}$ }
\newc{\omomnospace}{$\Omega_{0m}$}
\newcommand{\nn}{\nonumber}
\newc{\ra}{\Rightarrow}
\newc{\baodv}{$\frac{D_V}{r_s}$ }
\newc{\baodvnospace}{$\frac{D_V}{r_s}$}
\newc{\baoda}{$\frac{D_A}{r_s}$ } 
\newc{\baodanospace}{$\frac{D_A}{r_s}$}
\newc{\baodh}{$\frac{D_H}{r_s}$ }
\newc{\baodhnospace}{$\frac{D_H}{r_s}$}
\newc{\orcid}[1]{\href{https://orcid.org/#1}{\textcolor[HTML]{A6CE39}{\aiOrcid}}}
\newc{\ga}[1]{\textcolor{green}{[{\bf George}: #1]}}
\newc{\lk}[1]{\textcolor{orange}{[{\bf Lavrentios}: #1]}}
\newc{\lp}[1]{\textcolor{red}{[{\bf Leandros}: #1]}}
\newc{\snc}[1]{\textcolor{blue}{[{\bf Savvas}: #1]}}

\title{On the isotropy of SnIa absolute magnitudes in the Pantheon+ and SH0ES samples}

\author{Leandros Perivolaropoulos}\email{leandros@uoi.gr}
\affiliation{Department of Physics, University of Ioannina, GR-45110, Ioannina, Greece}

\date{\today}

\begin{abstract}
We use the hemisphere comparison method to test the isotropy of the SnIa absolute magnitudes of the Pantheon+ and SH0ES samples in various redshift/distance bins. We compare the identified levels of anisotropy in each bin with Monte-Carlo simulations of corresponding isotropised data to estimate the frequency of such levels of anisotropy in the context of an underlying isotropic cosmological. We find that the identified levels of anisotropy in all bins are consistent with the Monte-Carlo isotropic simulated samples. However, in the real samples for both the Pantheon+ and the SH0ES cases we find sharp changes of the level of anisotropy occuring at distances less than $40Mpc$. For the Pantheon+ sample we find that the redshift bin $[0.005,0.01]$ is significantly more anisotropic than the other 5 redshift bins considered. For the SH0ES sample we find a sharp drop of the anisotropy level at distances larger than about $30Mpc$. These anisotropy transitions are relatively rare in the Monte-Carlo isotropic simulated data and occur in $2\%$ of the SH0ES simulated data and at about $7\%$ of the Pantheon+ isotropic simulated samples. This effect is consistent with the experience of an off center observer in a $30Mpc$ bubble of distinct physics or systematics.
\end{abstract}
\maketitle

\section{Introduction}
\label{sec:Introduction}

The {\it cosmological principle} (CP)\cite{Aluri:2022hzs} is a cornerstone of the standard cosmological model \lcdm. It assumes that the Universe is homogeneous and isotropic on scales larger than about $100Mpc$. This assumption is consistent with most cosmological observations and allows the use of the FRW metric as a background where cosmological perturbation grow to form the observed large scale structure. The main cosmological observation that supports the CP is the isotropy of the Cosmic Microwave Background (CMB) and observations of the distribution of galaxies on scales larger than $100Mpc$ which are consistent with the onset of homogeneity and isotropy on these scales.

Despite of the simplicity and overall observational consistency of the CP, a few challenges have developed during the past several years which appear to consistently question the validity of the CP on cosmological scales larger than $100Mpc$ and motivate further tests to be imposed on its validity. Some of these challenges pointing to preferred directions on large cosmological scales include the quasar dipole\cite{Secrest:2020has,Zhao:2021fcp,Hu:2020mzd,Guandalin:2022tyl,Dam:2022wwh}, the radio galaxy dipole\cite{Wagenveld:2023key,Qiang:2019zrs,Singal:2023wni}, the bulk velocity flow\cite{Watkins:2023rll,Watkins:2014zaa,Wiltshire:2012uh,Nadolny:2021hti}, the dark velocity flow\cite{Atrio-Barandela:2014nda}, the CMB anomalies\cite{Copi:2010na,Schwarz:2015cma}, the galaxy spin alignment\cite{Tempel:2013gqa,Simonte:2023pzs,Shamir:2022obm}, the galaxy cluster anisotropies \cite{Bengaly:2016amk},
and a possible SnIa dipole\cite{Sorrenti:2022zat,McConville:2023xav}. These observational challenges of the CP may be indirectly connected with other tensions of the standard \lcdm model including the Hubble \cite{Perivolaropoulos:2021jda,Peebles:2022akh,Abdalla:2022yfr} and growth tensions \cite{Heymans:2020gsg,Troxel:2018qll,Asgari:2019fkq,vanUitert:2017ieu} where the best fit parameter values of the model $H_0$ and $\sigma_8$ appear to be inconsistent when probed by different observational data. These tensions may indicate that a new degree of freedom\cite{Sola:2016ecz,Anand:2017wsj,DiValentino:2021izs,DiValentino:2017zyq,Wang:2020dsc,Gomez-Valent:2017idt,DiValentino:2018gcu} is required to be introduced in the \lcdm model. For example it would be in principle possible that these tensions may disappear if the new degree of freedom corresponding to anisotropy was allowed to be introduced in the data analysis\cite{Tsagas:2021tqa,Asvesta:2022fts,Tsagas:2021ldz,Tsagas:2009nh,Colgain:2019pck,Krishnan:2021dyb,McConville:2023xav}.

An efficient method to test the CP is the use of type Ia supernovae (SnIa) which can map the expansion rate of the Universe up to redshifts of about 2.5. The latest and most extensive SnIa sample is the Pantheon+ sample\cite{Brout:2022vxf,Scolnic:2021amr,Brout:2021mpj}.It  provides equatorial coordinates, apparent magnitudes, distance moduli and other SnIa properties, derived from 1701 light curves of 1550
SnIa in a redshift range $z\in[0.001,2.26]$ compiled across 18 different surveys. This sample is  significantly improved over the first Pantheon sample of 1048 SnIa \cite{Pan-STARRS1:2017jku}, particularly at low redshifts $z$. 

The Pantheon+ sample has been used extensively for testing cosmological models and fitting cosmic expansion history parametrizations of $H(z)$ \cite{Brout:2022vxf}. It has also been used to identify and constrain possible velocity dipoles and compare with the corresponding velocity dipole obtained from the CMB. It was found that even though the observer velocity amplitude 
agrees with the dipole found in the cosmic microwave background, its direction is different at high significance\cite{Sorrenti:2022zat}. These results are in some tension with previous studies based on the previous Pantheon sample\cite{Horstmann:2021jjg}. The isotropy of $H_0$ has also been investigated using the hemisphere comparizon method \cite{Zhai:2022zif,McConville:2023xav,Krishnan:2021jmh,Krishnan:2021dyb,Krishnan:2022uar,Luongo:2021nqh,Zhai:2023ubp} and relatively small but statistically significant anisotropy level was identified in the direction of the CMB dipole. Since there is degeneracy between $H_0$ and the SnIa absolute magnitude a possible anisotropy in the best fit value of $H_0$ is probably connected with an anisotropy of the SnIa absolute magnitudes.

Recent analyses have pointed out that a sudden change (transition) of the SnIa absolute magnitude by about $0.2$ at a transition redshift $z\lesssim 0.01$ could imply a lower value of $H_0$ compared to the one measured in the context of a standard distance ladder approach that does not incorporate this transition degree of freedom \cite{Marra:2021fvf,Alestas:2020zol,Perivolaropoulos:2021jda,Perivolaropoulos:2022khd}. Such a sudden change could occur in the context of a  gravitational physics transition taking place globally during the past 150 Myrs, or locally in a bubble of scale of about $40Mpc$ and shifting the strength of gravity by a few percent. Such a transition would lead to a systematic dimming of SnIa which could be incorrectly interpreted as a higher value of $H_0$ due to the degeneracy between $H_0$ and $M$. It has recently been shown that such a scenario is consistent\cite{Perivolaropoulos:2022khd} with the SH0ES data\cite{Riess:2021jrx}, with the Pantheon+ data\cite{Perivolaropoulos:2023iqj} and with other astrophysical and geological observations\cite{Alestas:2021nmi,Perivolaropoulos:2022vql}. 

\begin{figure}[ht!]
\centering
\includegraphics[trim=0 100 0 0, clip, width=0.47\textwidth]{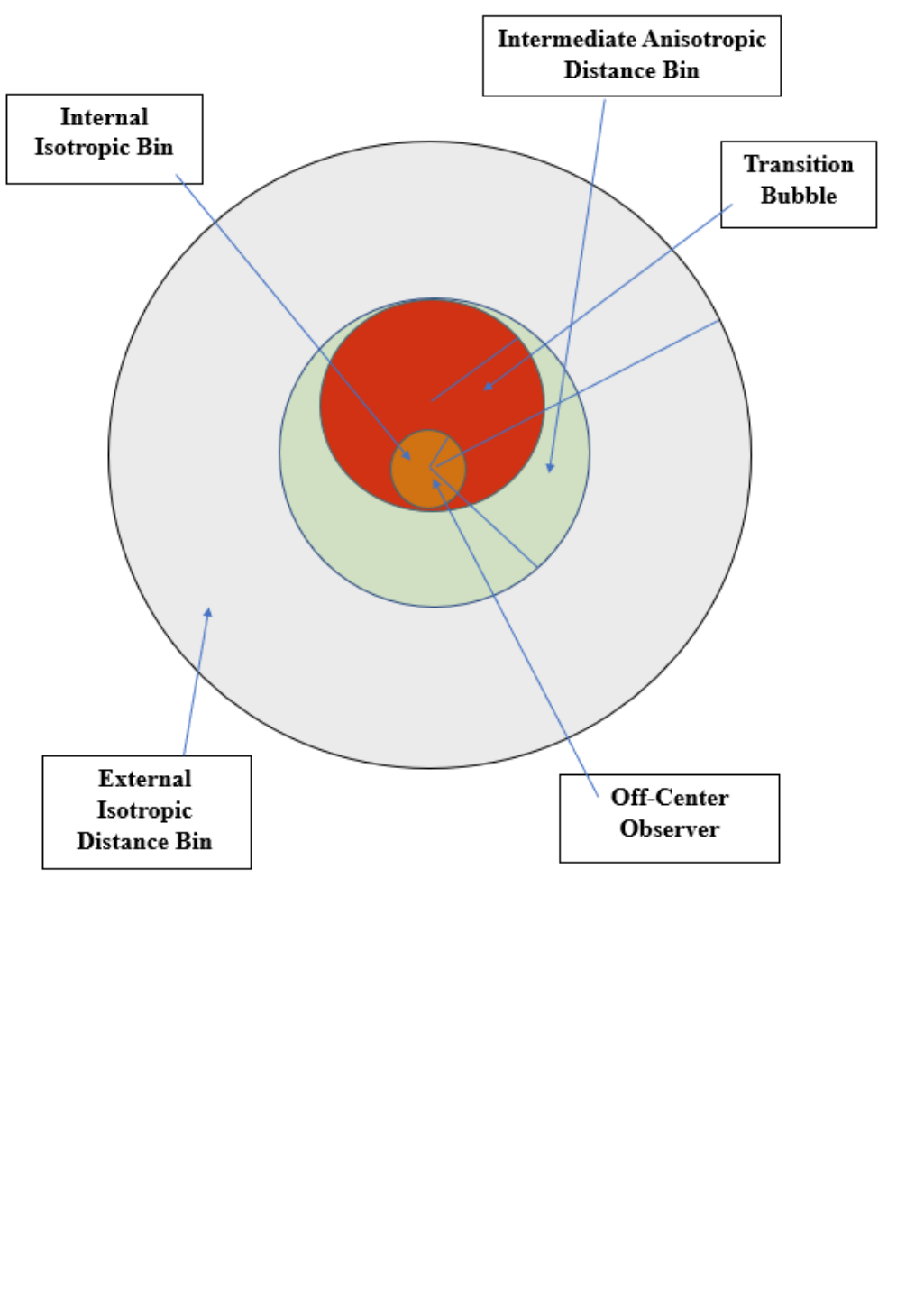}
\caption{Demonstration of the predicted change of the anisotropy level of SnIa absolute magnitudes experienced by an off-center observer (center of small pink circle) in a gravitational transition bubble (red circle) where the gravitational constant is different by a few percent. A low anisotropy is expected in the small pink distance bin, a large anisotropy level is expected for the distance bin between the pink inner circle and the green outer circle and finally a smaller anisotropy level is expected for the distance bin between the green and the grey outer circles.}
\label{fig:aniso-obs1}
\end{figure}

This scenario may be realized in the context of a local $40Mpc$ bubble rather than a global transition at late times. In this case an off-center observer would detect a sudden increase of the anisotropy of the SnIa absolute magnitudes in spherical distance bins that include parts of both the inside and the outside region of the transition bubble (intermediate anisotropic distance bin in Fig. \ref{fig:aniso-obs1}). Distance bins that are much larger of much smaller than the radius of the transition bubble would appear isotropic to the off center observer as shown in Fig. \ref{fig:aniso-obs1}.

In the context of this prediction of the transition model for the anisotropy of the SnIa absolute magnitudes, the following questions arise:
\begin{itemize}
\item 
What is an efficient and general purpose statistic to quantitatively describe the anisotropy level of the SnIa absolute magnitudes of the Pantheon+ sample?
\item 
Given such a statistic, what is the level of anisotropy of the SnIa absolute magnitudes for various distance (redshift) bins of the Patheon+ and SH0ES samples?
\item 
What are the directions corresponding to these anisotropies for each redshift bin and how do these directions relate to the CMB dipole?
\item
What are the corresponding results expected in the context of Monte-Carlo simulated data of Pantheon+ in the context of isotropy?
\item 
Are the above answers for the Pantheon+ and SH0ES samples consistent with each other? 
\end{itemize}
The goal of the present analysis is to address these questions. In order to address these questions we obtain the SnIa absolute magnitudes as
\be
M=m_B-\mu
\label{absmagndef}
\ee
where $m_B$ is the apparent magnitude of SnIa and $\mu$ is the distance modulus obtained either from the Cepheids co-hosted with the SnIa (published with the Pantheon+ data) or from the best fit \lcdm model as obtained from Pantheon+.

The structure of this paper is the following: In the next section \ref{II} we describe the statistic $\Sigma_{max}$ used to quantify the anisotropy level in our analysis and the hemisphere comparison method implemented in the context of this statistic. The implementation of this method is presented in section \ref{III} for both the Pantheon+ and the SH0ES samples. The results of this anisotropy analysis are presented and qualitatively compared with those expected in the context of the local transition model predictions. Finally in section \ref{IV} we summarize our main results, discuss their implications and describe possible future extensions of this analysis.

\section{The hemisphere comparison method}
\label{II}
The hemisphere comparison method\cite{Schwarz:2007wf,Antoniou:2010gw} along with the dipole method has been extensively used\cite{Mariano:2012ia,Mariano:2012wx,Deng:2018yhb,Chang:2014nca,Kazantzidis:2020tko,Deng:2018jrp} for the identification of the level of anisotropy and the corresponding direction in a wide range of cosmological data. The implementation of this method in the present analysis involves the following steps:
\begin{figure*}[ht!]
\centering
\includegraphics[width=\textwidth]{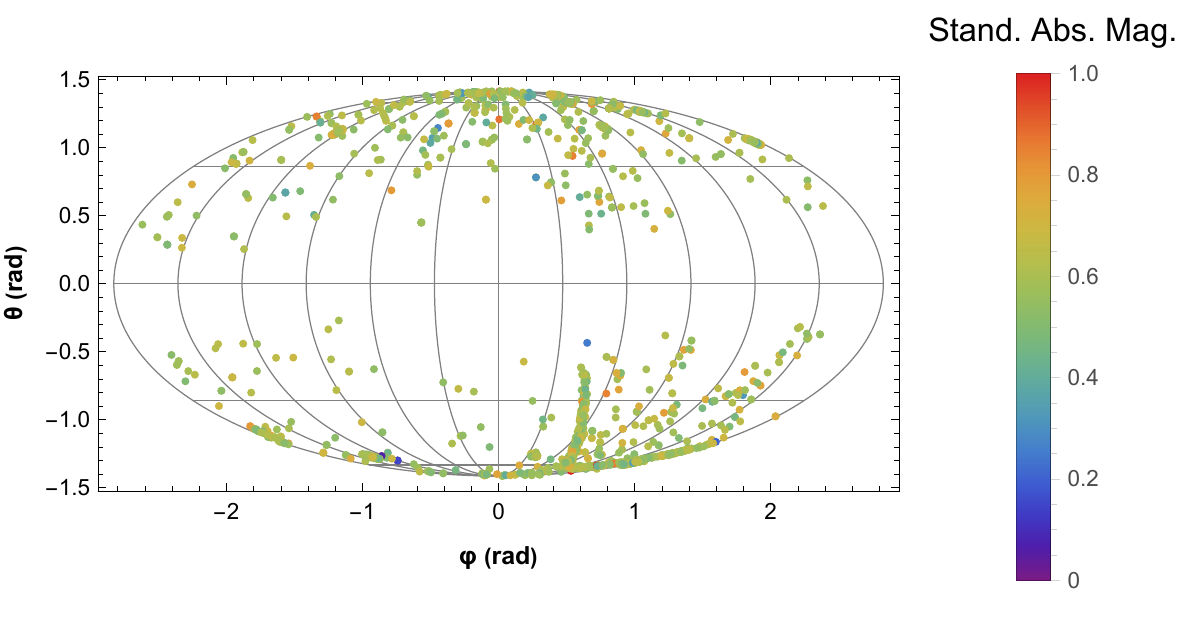}
\caption{Mollweide projection in the galactic frame of the sky distribution of the Pantheon+ SnIa. Both angular coordinates are in radians and the azimouthal coordinate ranges in the projected range of $[-\pi,\pi]$ The color describes the SnIa standardized absolute magnitudes $\bar M$. }
\label{fig:molwfullpanth}
\end{figure*}

\begin{figure}[ht!]
\centering
\begin{subfigure}{0.5\textwidth}
\includegraphics[width=\textwidth]{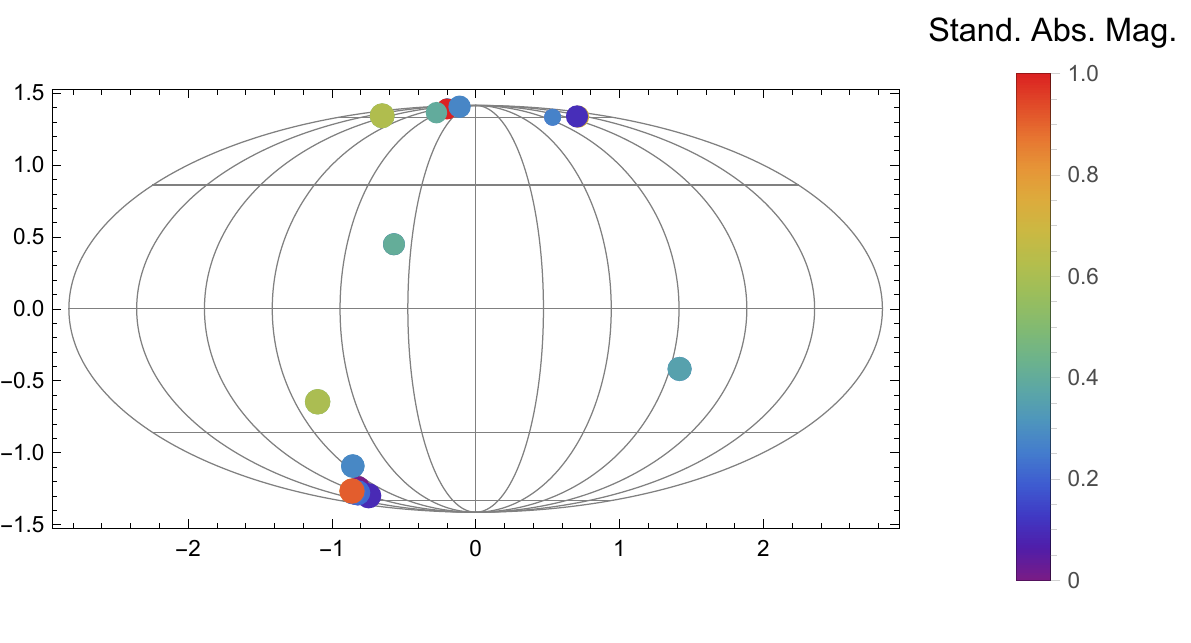}
\caption{The SnIa distribution in the lowest redshift bin $[0.001,0.005]$. The color denotes the standardized absolute magnitude $\bar M$ and the radius of each point increases with redshift. There are 29 SnIa but not all of them are shown due to overlap of directions since in this distance range there are SnIa that share the same host and thus the same galactic coordinates.}
\label{fig:molwpanth0001-0005}
\end{subfigure}
\hfill
\begin{subfigure}{0.5\textwidth}
\includegraphics[width=\textwidth]{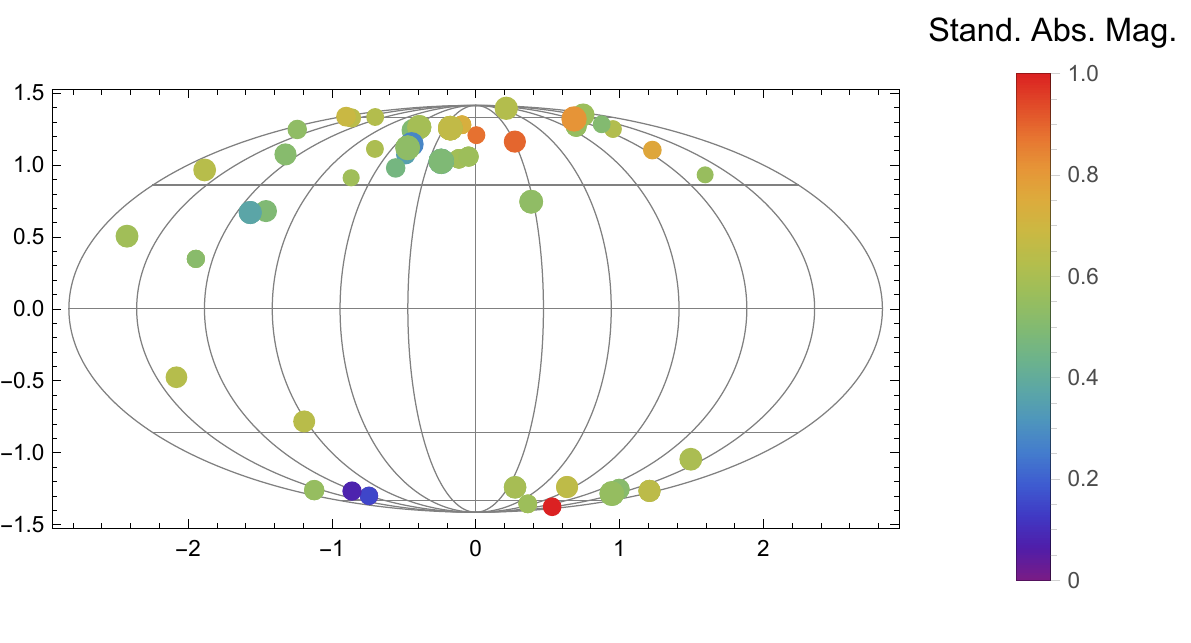}
\caption{The SnIa distribution in the  redshift bin $[0.005,0.01]$.}
\label{fig:molwpanth0005-001}
\end{subfigure}
\caption{The SnIa sky distribution in the two lowest redshift bins.}
\label{fig:molpanthall}
\end{figure}

\begin{itemize}
\item 
Use the publicly available Pantheon+ data\cite{Brout:2022vxf} to estimate the absolute magnitude of each one of the 1701 SnIa using eq, (\ref{absmagndef}) where $\mu$ is obtained from the Cepheid estimate for SnIa in Cepheid hosts or from the best fit Pantheon+ \lcdm model with $\Omega_{0m}=0.333$, $h=0.734$ for the rest of the SnIa\cite{Brout:2022vxf,Perivolaropoulos:2023iqj}. For these SnIa we set
\be
\mu(z)=5\; log_{10}(d_L(z;\Omega_{0m},h)/Mpc)+25
\label{mumodel}
\ee
where 
\be
d_L(z;\Omega_{0m},h)=c(1+z)\int_0^z \frac{dz'}{H(z';\Omega_{0m},h)}
\label{dlz}
\ee
is the luminosity distance and 
\be
H(z)=100 \; h \left[\Omega_{0m} (1+z)^2 +(1-\Omega_{0m})\right]^{1/2}
\label{hzlcdm}
\ee
is the Hubble expansion rate for \lcdm ($H_0=100\; h\; km/sec\cdot Mpc$). The residual absolute magnitude may be defined as
\be
M_{res}\equiv M_i-M_{SH0ES}
\label{mresdef}
\ee
where $M_{SH0ES}=-19.2483$ is the best fit SnIa absolute magnitude as obtained from the SH0ES data\cite{Riess:2021jrx}. In what follows, we use the residual absolute magnitudes $M_{res}$ but omit the index $_{res}$ for simplicity. The uncertainty of each (residual) $M$ may be estimated using the uncertainties of $m$ and $\mu$ from eq. (\ref{absmagndef}) as 
\be
\sigma_M=\sqrt{\sigma_m^2+\sigma_\mu^2}
\label{sigmam}
\ee
where $\sigma_m$ and $\sigma_\mu$ are uncertainties of the apparent magnitude  and the distance modulus as provided by the Pantheon+ sample\footnote{The distance modulus provided in the Pantheon+ data assumes that the absolute magnitude is the same for all SnIa and is equal to $M_{SH0ES}$.}. The mean of these residual absolute magnitudes is 0. For plotting convenience we also use the standardized absolute magnitudes defined as
\be
{\bar M}\equiv \frac{M-M_{min}}{M_{max}-M_{min}}
\label{mstand}
\ee
where $M_{max}$, $M_{min}$ are the maximum and minimum values of the SnIa absolute magnitudes. Obviously by definition ${\bar M}\in [0,1]$. 
\item 
The published SnIa coordinates of SnIa in Pantheon+ are equatorial coordinates (RA,DEC). We thus convert them to galactic coordinates $(l,b)$ using standard algorithms\cite{DuffettSmithZwart2011}. The resulting map of the SnIa absolute magnitudes is shown in Fig. \ref{fig:molwfullpanth}. In Fig. \ref{fig:molpanthall} we also show the corresponding angular distribution for two redshift bins: $z\in [0.001,0.005]$ in Fig. \ref{fig:molwpanth0001-0005}  (29 points, some of them overlapping) and $z\in [0.005,0.01]$ in Fig. \ref{fig:molwpanth0005-001} (82 points). The color of the points denotes the standardized absolute magnitude while the radius of the points in Fig. \ref{fig:molpanthall} increases with the redshift.
\item 
In a given redshift bin we consider a random direction in galactic coordinates and split the sample in two subsamples: one with SnIa in the 'North' hemisphere with respect to this random direction and one in the 'South' hemisphere. For each hemisphere we construct the weighted average (residual) absolute magnitude ($M_N$ and $M_S$) and the corresponding uncertainties ($\sigma_N$ and $\sigma_S$) using the equations
\be
M_N=\frac{\sum_i{M_{N,i}/\sigma_{N,i}^2}}{\sum_i{1/\sigma_{N,i}^2}}
\label{mnave}
\ee
where the sum runs over the SnIa of the 'North' hemisphere. Similarly we obtain $M_S$ for the 'South' hemisphere with respect to the given random direction. The corresponding uncertainty of $M_N$ is obtained as 
\be
\sigma_N=\sqrt{\left(\sum_i{1/\sigma_{N,i}^2}\right)^{-1}}
\label{sigmn}
\ee
and similarly for $\sigma_S$.
\item 
We construct the statistical variable representing the anisotropy level for the given random direction 
\be
\Sigma\equiv \frac{\vert M_N - M_S \vert}{\sqrt{\sigma_N^2 + \sigma_S^2}}
\label{sigmadef}
\ee
\item 
We evaluate $\Sigma$ for 1000 random directions and select the maximum anisotropy direction leading to the maximum anisotropy level $\Sigma_{max}$. 
\item 
We generate Monte-Carlo simulated versions of the Pantheon+ and SH0ES residual absolute magnitudes by replacing each SnIa absolute magnitude my a random value selected from a Gaussian distribution with 0 mean and standard deviation given by eq. (\ref{sigmam}) for each residual absolute magnitude $M$.
\item 
For each one of the Monte-Carlo Pantheon+ absolute magnitude samples we identify $\Sigma_{max}$ and thus we find the mean $\Sigma_{max}$ and its standard deviation as obtained from 30 Monte-Carlo Pantheon+ or SH0ES realizations. We thus compare the anisotropy level $\Sigma_{max}$ of the real data with the anticipated $1\sigma$ range of $\Sigma_{max}$ obtained from the Monte-Carlo simulated data for various redshift bins and for the full samples data. If the value of $\Sigma_{max}$ from the real data is larger than the $1\sigma$ Monte Carlo range this could be interpreted as a hint for statistically significant anisotropy in the given redshift bin. In the opposite case of $\Sigma_{max}$ less than the Monte-Carlo range it could be interpreted as an overestimation of the uncertainties that generated the Monte-Carlo  samples.
\end{itemize}

In the next section we implement the above described hemisphere comparison method on both the Pantheon+ and the SH0ES maps of the SnIa residual absolute magnitudes to identify possible abnormal levels of anisotropy on various redshift bins and/or abnormal variations of the anisotropy levels among different redshift bins.

\section{Isotropy of SnIa absolute magnitudes}
\label{III}
\subsection{Isotropy of Pantheon+ SnIa Sample}

\begin{widetext}
\begin{table*}[ht]
\centering
\begin{tabular}{|c|c|c|c|c|c|c|c|}
\hline
\textbf{Bin} & \textbf{$z_{min}$} & \textbf{$z_{max}$} & \textbf{$\Sigma_{max}$} & \textbf{Monte-Carlo Range} & \textbf{Data in Bin} & \textbf{Anisotropy Axis $ (l,b)$} & \textbf{$\theta_{CMB}$} \\
\hline
1 & 0.001 & 0.005 & 0.51 & $2.04 \pm 0.45$ & 29 & ($222.54^\circ, 35.85^\circ$) & $32.68^\circ$ \\
2 & 0.005 & 0.01 & 1.93 & $1.93 \pm 0.28$ & 82 & $ (100.44^\circ, 33.51^\circ)$  & $97.17^\circ$ \\
3 & 0.01 & 0.05 & 1.70 & $2.71 \pm 0.71$ & 534 & ($331.58^\circ, 33.71^\circ$) & $51.34^\circ$ \\
4 & 0.05 & 0.10 & 1.32 & $2.36 \pm 0.45$ & 96 & ($85.51^\circ, 11.75^\circ$) & $120.23^\circ$\\
5 & 0.10 & 0.30 & 1.10 & $2.19 \pm 0.74$ & 466 & ($188.04^\circ, 4.63^\circ$) & $77.18^\circ$ \\
6 & 0.30 & 2.30 & 1.16 & $2.24 \pm 0.40$ & 494 & ($241.58^\circ, 18.89^\circ$) & $34.33^\circ$ \\
All & 0.001 & 2.30 & 1.50 & $2.30 \pm 0.70$ & 1701 & ($136.39^\circ, 16.12^\circ$) & $100.72^\circ$ \\
\hline
\end{tabular}
\caption{The anisotropy level $\Sigma_{max}$ in six redshift bins of the Pantheon+ data and the corresponding anisotropy directions in galactic coordinates. The anticipated $1\sigma$ range from Monte-Carlo simulations is also shown and the angle of each anisotropy direction with the CMB dipole is shown in the last column.}
\label{tab:smax}
\end{table*}
\end{widetext}

We split the Pantheon+ data is six redshift bins and for each redshift bin we identify the maximum anisotropy level $\Sigma_{max}$ and the corresponding direction axis in galactic coordinates as well as its angle with the CMB dipole which is towards $(l,b)=(264^\circ,48^\circ)$ in galactic coordinates. Since $\Sigma_{max}$ is positive definite, the anisotropy axis has no preferred sign and we define it to point towards the north galactic hemisphere direction. For each bin we also identify the anticipated $1\sigma$ range of $\Sigma_{max}$ obtained from 30 Monte-Carlo realizations.
The results are shown in Table \ref{tab:smax}

\begin{figure}[ht!]
\centering
\includegraphics[width = 0.47 \textwidth]{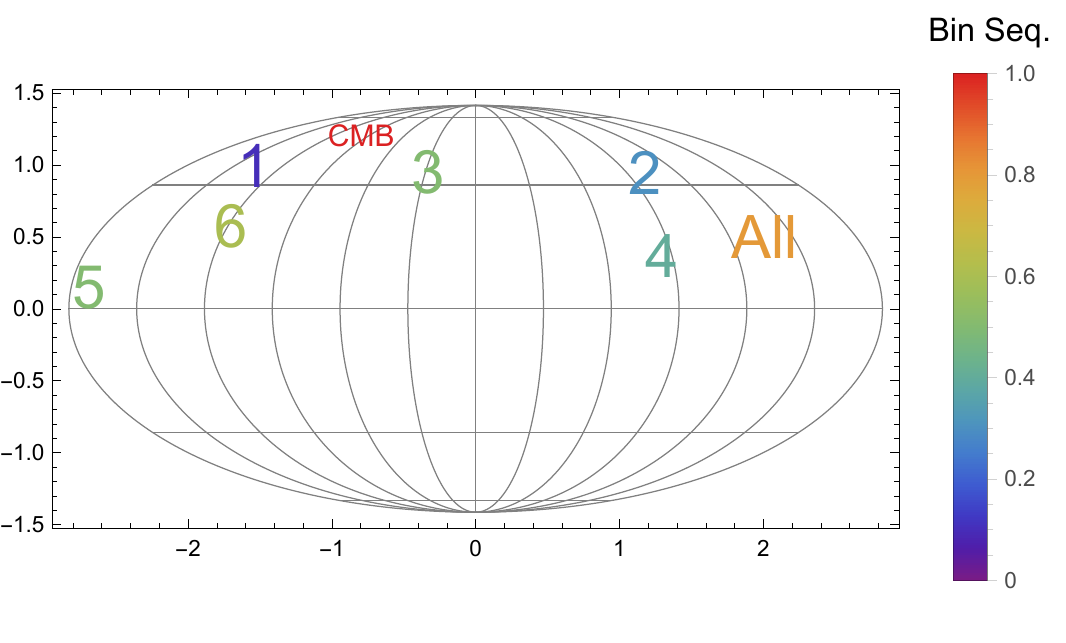}
\caption{the maximum anisotropy directions for each one of the six considered redshift bins shown in Table \ref{tab:smax} and in Fig. \ref{fig:smaxbinpanth}.}
\label{fig:plbindirsalln}
\end{figure}

The corresponding anisotropy directions for each redshift bin are shown in Fig. \ref{fig:plbindirsalln}. Notice that even though the bins 1, 3 and 6 are close to the CMB dipole direction there is no overall trend of the full dataset or for all bins to be correlated with the CMB data. 

The Mollweide projection is in the galactic coordinate frame but the azimouthal angle ranges in $[-\pi,\pi]$ in radians, in contrast to the galactic coordinate $l$ which ranges in $[0^\circ,360^\circ]$. Thus the CMB dipole direction appears in the left part of the North galactic hemisphere. Also both angular coordinates are shown in radians.

\begin{figure}[ht!]
\centering
\includegraphics[width = 0.47 \textwidth]{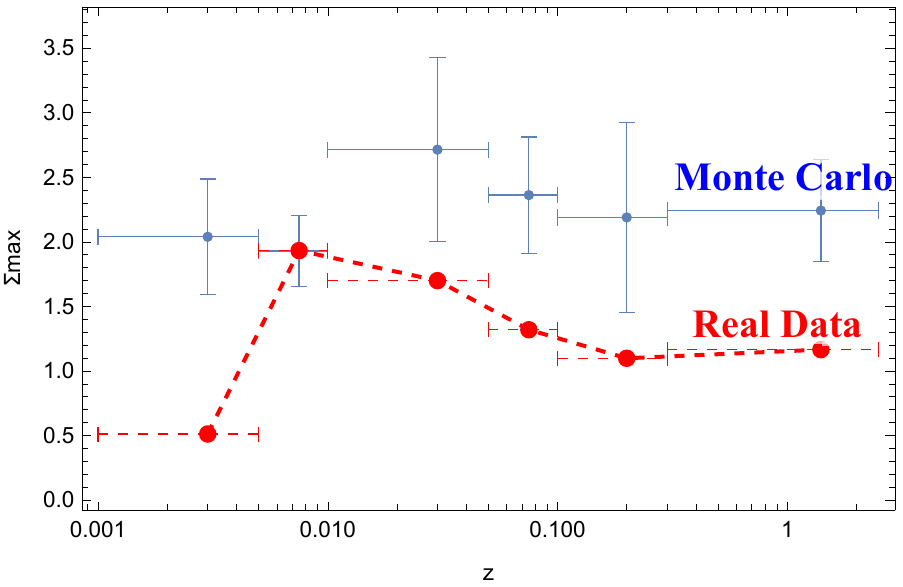}
\caption{The value of the anisotropy level $\Sigma_{max}$ for the Pantheon+ data (red line and points) in six redshift bins. The corresponding ranges of the isotropic Monte-Carlo data are also shown (blue points). }
\label{fig:smaxbinpanth}
\end{figure}  

The maximum anisotropy levels $\Sigma_{max}$ for each redshift bin along with the corresponding Monte-Carlo $1\sigma$ ranges are shown in Fig. \ref{fig:smaxbinpanth}. There are two points to notice in this Figure. First, there is a clear maximum in the level of anisotropy is the bin $z\in [0.005,0.01]$ (distance range $d\in [20,40]Mpc$) which is not anticipated from the Monte-Carlo ranges. This is consistent with the prediction of the off-center observer hypothesis discussed in the Introduction section \ref{sec:Introduction} (see Fig. \ref{fig:aniso-obs1}). Second, the real data appear to have systematically lower level of anisotropy in most bins than the level anticipated on the basis of Monte-Carlo simulations. This could be due to an overestimate of the uncertainties as obtained from eq. (\ref{sigmam}) and the fact that the construction of the Monte-Carlo simulations has not taken into account possible correlations among the different absolute magnitudes expressed through the covariance matrix.

\begin{figure}[ht!]
\centering
\includegraphics[width = 0.47 \textwidth]{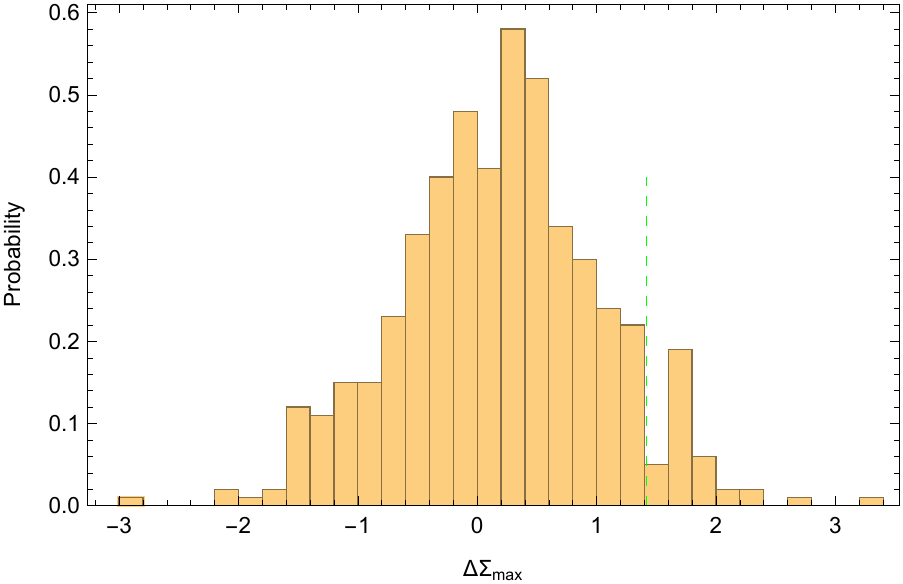}
\caption{Probability distribution of the change of level of anisotropy between the lowest two redshift bins as obtained from the Monte-Carlo simulations. The real data value of $\Delta \Sigma_{max}$ is denoted by the dashed green line.}
\label{fig:plprobdsmax}
\end{figure}

In order to estimate the statistical significance of the observed shift of the anisotropy level in the first two redshift bins, in Fig. \ref{fig:plprobdsmax}, we have constructed a histogram of the probability of change of $\Sigma_{max}$ in the first two redshift bins obtained from the Monte-Carlo simulations. The dashed green line corresponds to the $\Delta \Sigma_{max}$ shift for the real data. About $7\%$ of the Monte-Carlo samples had a $\Delta \Sigma_{max}$ shift equal or larger than that of the real data. This probability would get smaller if we had also demanded from the Monte-Carlo data to have a smaller $\Sigma_{max}$ on higher redshift bins as observed in the real data and as predicted in the context of the off-center observer hypothesis.    

\subsection{Isotropy of SH0ES SnIa Sample}

\begin{figure}[ht!]
\centering
\includegraphics[width = 0.47 \textwidth]{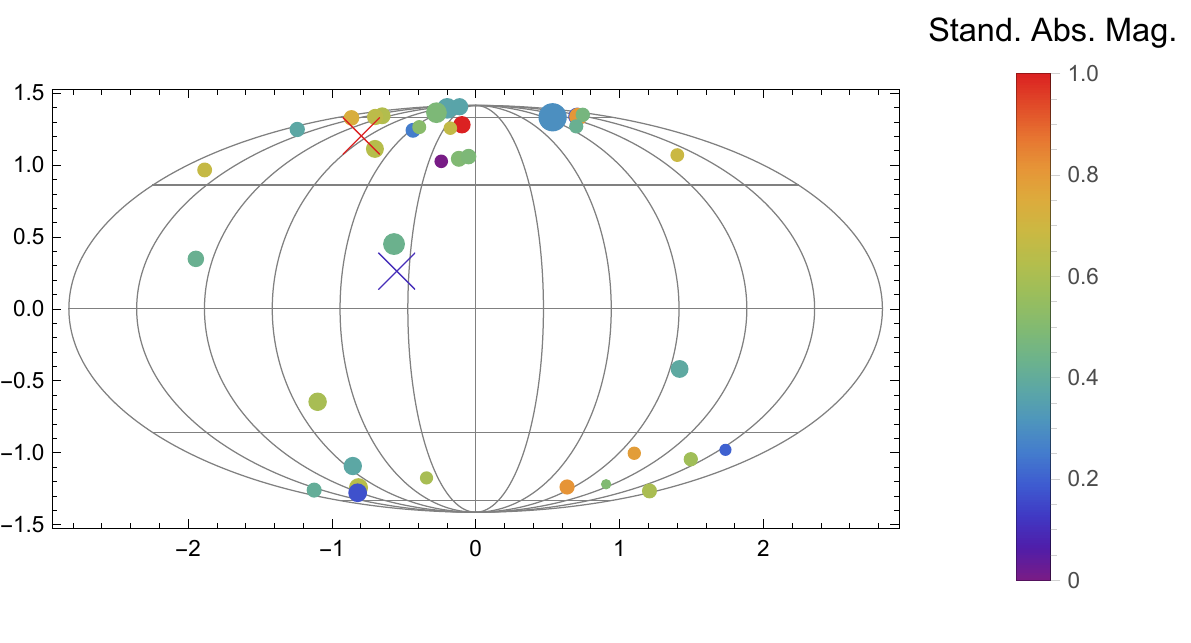}
\caption{The angular distribution in the sky of 37 SnIa host absolute magnitudes obtained as weighted averages of the SnIa absolute magnitudes in each host. The CMB dipole direction is shown with the red cross and the maximum anisotropy axis is shown with the blue cross. Mild alignment is observed as was the case for the low distance bins of the Pantheon+ sample (see Fig. \ref{fig:plbindirsalln}). }
\label{fig:plsh0esmolw}
\end{figure}

The identified change of anisotropy level identified for the second distance bin $[20,40]Mpc$ in the Pantheon+ data is also identified in the SH0ES data of SnIa hosts hosts shown in Fig. \ref{fig:plsh0esmolw}. 
For this part of the analysis we have considered the weighted average absolute magnitude of each one of the 37 Cepheid hosts of the SH0ES data which are very weakly correlated. These absolute magnitudes and their derivation are discussed in detail in Ref. \cite{Perivolaropoulos:2023iqj}. The corresponding SH0ES data are shown in Table \ref{tab:SH0ES}.

\begin{figure}[ht!]
\centering
\includegraphics[width = 0.47 \textwidth]{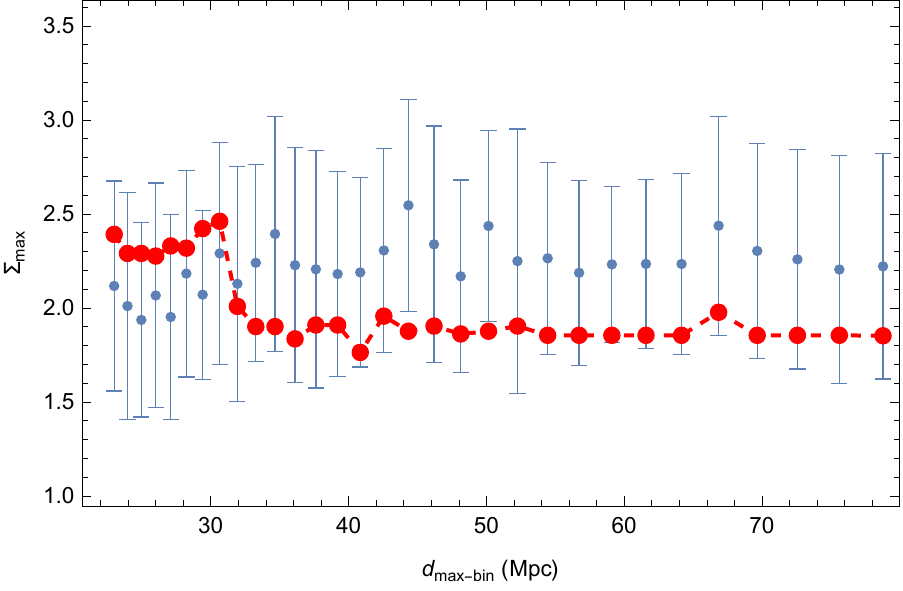}
\caption{The anisotropy level for cumulative distance bins with maximum distance $d_{max}$ as obtained from the distance modulus of each host.}
\label{fig:plsh0essmaxbin}
\end{figure}

Due to the small number of hosts in the SH0ES data we have considered cumulative distance bins from low to high distances. The maximum distance $d_{max}$ of each cumulative bin was obtained from the largest distance modulus of the bin as
\be d_{max}=10^\frac{\mu_{Ceph,max}-25}{5}
\label{dmaxmumax}
\ee
where $\mu_{Ceph,max}$ maximum distance modulus of each cumulative distance bin as obtained from Cepheids.
Thus each cumulative bin which includes hosts with distances in the range $d\in [d_{min},d_{max}]$ where $d_{min}=6.8Mpc$ corresponding to the closest SnIa+Cepheid host (M101) and $d_{max}$ is the maximum distance of the bin. In each distance bin we implement the hemisphere comparison method and identify the maximum anisotropy level. Thus we plot $\Sigma_{max}$ in terms of $d_{max}$ in Fig. \ref{fig:plsh0essmaxbin}. In terms of alignment of the maximum anisotropy direction with the CMB dipole  we find mild alignment with the CMB dipole for the full SH0ES data. Note however that there are large uncertainties in terms of determining the anisotropy direction due to the small number of SnIa in the SH0ES data.

The anisotropy level for the SH0ES data shown in Fig. \ref{fig:plsh0essmaxbin} is consistent with the corresponding results for the Pantheon+ sample.  It shows a rapid decrease of the anisotropy level at $d_{max}\simeq 30Mpc$ even though the anisotropy level for all cumulative bins are consistent with the $1\sigma$ Monte-Carlo ranges shown also in Fig. \ref{fig:plsh0essmaxbin}. Due to the cumulative nature of these bins and the small number of datapoints, the lowest distance bin of the Pantheon+ analysis is not probed and thus the rapid increase of the anisotropy level is not manifest in the SH0ES data case even though the sharp decrease of the anisotropy level is clearly evident. The SH0ES data that were used for the construction of Figs \ref{fig:plsh0esmolw} and \ref{fig:plsh0essmaxbin} are shown in Table \ref{tab:SH0ES}.

\begin{table*}[ht]
\centering
\begin{tabular}{|c|c|c|c|c|c|c|c|c|c|}
\hline
\textbf{Host name} & \textbf{SnIa name} & \textbf{$\mu_{Ceph}$} & \textbf{$\sigma_{\mu_{Ceph}}$} & \textbf{$m_B$} & \textbf{$\sigma_{m_B}$} & \textbf{M} & \textbf{$\sigma_M$} & \textbf{RA} & \textbf{DEC} \\
\hline
M101 & 2011fe & 29.178 & 0.041 & 9.78 & 0.115 & -19.398 & 0.122 & 210.774 & 54.274 \\
N5643 & 2017cbv & 30.546 & 0.052 & 11.229 & 0.054 & -19.317 & 0.075 & 218.143 & -44.134 \\
N4424 & 2012cg & 30.844 & 0.128 & 11.487 & 0.192 & -19.357 & 0.230 & 186.803 & 9.420 \\
N4536 & 1981B & 30.835 & 0.05 & 11.551 & 0.133 & -19.284 & 0.142 & 188.623 & 2.199 \\
N1448 & 2021pit & 31.287 & 0.037 & 12.095 & 0.112 & -19.191 & 0.118 & 56.125 & -44.632 \\
N1365 & 2012fr & 31.378 & 0.056 & 11.9 & 0.092 & -19.478 & 0.107 & 53.4 & -36.127 \\
N1559 & 2005df & 31.491 & 0.061 & 12.141 & 0.086 & -19.35 & 0.105 & 64.407 & -62.770 \\
N2442 & 2015 & 31.45 & 0.064 & 12.234 & 0.082 & -19.216 & 0.104 & 114.066 & -69.506 \\
N7250 & 2013dy & 31.628 & 0.125 & 12.283 & 0.178 & -19.345 & 0.217 & 334.573 & 40.570 \\
N3982 & 1998aq & 31.722 & 0.071 & 12.252 & 0.078 & -19.47 & 0.105 & 179.108 & 55.129 \\
N4038 & 2007sr & 31.603 & 0.116 & 12.409 & 0.106 & -19.194 & 0.157 & 180.47 & -18.973 \\
N3972 & 2011by & 31.635 & 0.089 & 12.548 & 0.094 & -19.087 & 0.129 & 178.94 & 55.326 \\
N4639 & 1990N & 31.812 & 0.084 & 12.454 & 0.124 & -19.358 & 0.149 & 190.736 & 13.257 \\
N5584 & 2007af & 31.772 & 0.052 & 12.804 & 0.079 & -18.968 & 0.094 & 215.588 & -0.394 \\
N3447 & 2012ht & 31.936 & 0.034 & 12.736 & 0.089 & -19.2 & 0.0953 & 163.345 & 16.776 \\
N2525 & 2018gv & 32.051 & 0.099 & 12.728 & 0.074 & -19.323 & 0.123 & 121.394 & -11.438 \\
N3370 & 1994ae & 32.12 & 0.051 & 12.937 & 0.082 & -19.183 & 0.0965 & 161.758 & 17.275 \\
N5861 & 2017erp & 32.223 & 0.099 & 12.945 & 0.107 & -19.278 & 0.146 & 227.312 & -11.334 \\
N5917 & 2005cf & 32.363 & 0.12 & 13.079 & 0.095 & -19.284 & 0.153 & 230.384 & -7.413 \\
N3254 & 2019np & 32.331 & 0.076 & 13.201 & 0.074 & -19.13 & 0.106 & 157.342 & 29.510 \\
N3021 & 1995al & 32.464 & 0.158 & 13.114 & 0.116 & -19.35 & 0.196 & 147.733 & 33.552 \\
N1309 & 2002fk & 32.541 & 0.059 & 13.209 & 0.082 & -19.332 & 0.101 & 50.523 & -15.400 \\
N4680 & 1997bp & 32.599 & 0.205 & 13.173 & 0.205 & -19.426 & 0.290 & 191.724 & -11.642 \\
N1015 & 2009ig & 32.563 & 0.074 & 13.35 & 0.094 & -19.213 & 0.119 & 39.548 & -1.312 \\
N7541 & 1998dh & 32.5 & 0.119 & 13.418 & 0.128 & -19.082 & 0.175 & 348.668 & 4.537 \\
N2608 & 2001bg & 32.612 & 0.154 & 13.443 & 0.166 & -19.169 & 0.226 & 128.829 & 28.468 \\
N3583 & 2015so & 32.804 & 0.08 & 13.509 & 0.093 & -19.295 & 0.123 & 168.654 & 48.318 \\
U9391 & 2003du & 32.848 & 0.067 & 13.525 & 0.084 & -19.323 & 0.107 & 218.649 & 59.334 \\
N0691 & 2005W & 32.830 & 0.109 & 13.602 & 0.139 & -19.228 & 0.177 & 27.690 & 21.759 \\
N5728 & 2009Y & 33.094 & 0.205 & 13.514 & 0.115 & -19.580 & 0.235 & 220.599 & -17.246 \\
M1337 & 2006D & 32.92 & 0.123 & 13.655 & 0.106 & -19.265 & 0.162 & 193.141 & -9.775 \\
N3147 & 2021hpr & 33.014 & 0.165 & 13.8536 & 0.093 & -19.160 & 0.190 & 154.161 & 73.400 \\
N5468 & 2002cr & 33.116 & 0.074 & 13.942 & 0.053 & -19.173 & 0.091 & 211.657 & -5.439 \\
N7329 & 2006bh & 33.246 & 0.117 & 14.03 & 0.079 & -19.216 & 0.141 & 340.067 & -66.485 \\
N7678 & 2002dp & 33.187 & 0.153 & 14.09 & 0.093 & -19.097 & 0.179 & 352.127 & 22.428 \\
N0976 & 1999dq & 33.709 & 0.149 & 14.25 & 0.103 & -19.459 & 0.181& 38.498 & 20.975 \\
N0105 & 2007A & 34.527 & 0.25 & 15.25 & 0.133 & -19.277 & 0.283 & 6.320 & 12.886 \\
\hline
\end{tabular}
\caption{The SH0ES data that were used for the construction of Figs \ref{fig:plsh0esmolw} and \ref{fig:plsh0essmaxbin}. For each host we show a representative SnIa that was used for determining the galactic coordinates of the host.}
\label{tab:SH0ES}
\end{table*}

\begin{figure}[ht!]
\centering
\includegraphics[width = 0.47 \textwidth]{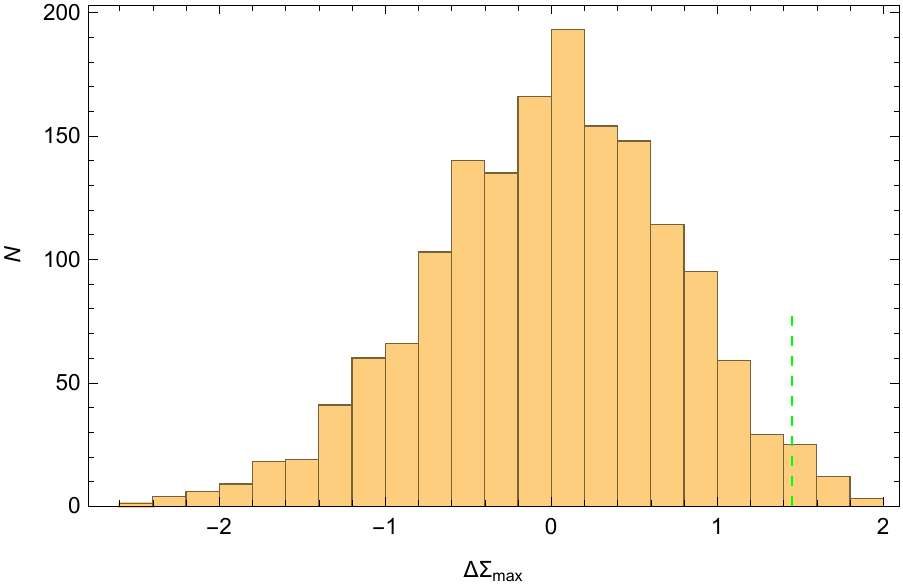}
\caption{Histogram describing the probability of finding a change $\Delta\Sigma_{max}$ of the maximum anisotropy level, between the low distance bin for the SH0ES data. The real data value of $\Delta \Sigma_{max}=1.45$ is denoted with the green dashed line. }
\label{fig:plhistoshoes}
\end{figure} 

In order to estimate the probability that the observed sharp decrease of the anisotropy level would occur in the context of generically isotropic data we have constructed $40$ Monte-Carlo simulated SH0ES host absolute magnitude samples for each one of two distance bins: a low distance bin with $d\in [d_{min},d_{min}+20]Mpc$ and a high distance bin with $d\in [d_{min}+20]Mpc,d_{max}]$ where $d_{min}=6.8Mpc$ is the minimum SnIa+Cepheid host distance of the SH0ES sample and $d_{max}=80Mpc$ is the corresponding maximum distance. We thus have $40\times 40=1600$ bin pairs. For each pair we find the maximum anisotropy level difference $\Delta \Sigma_{max}$ and plot a histogram for the frequency of these differences in Fig. \ref{fig:plhistoshoes}. For the real data we have $\Delta \Sigma_{max}=1.45$ shown with a green dashed line in Fig. \ref{fig:plhistoshoes}. Only 32 of the 1600 bin pairs had a difference $\Delta \Sigma_{max}\geq 32$ ie larger than or equal to $\Delta \Sigma_{max}$ of the real data. This corresponds to a probability of about $2\%$. Thus the anisotropy level transition in the SH0ES data is fairly significant statistically (almost at $3\sigma$.

\section{Discussion-Conclusion}
\label{IV}

We have used the hemisphere comparison method to investigate the level of SnIa absolute magnitude anisotropy in the Pantheon+ and SH0ES samples. Our analysis is distinct from some previous studies \cite{Sorrenti:2022zat} in that we have not focused on the velocity dipole but on intrinsic properties of SnIa corresponding to their absolute magnitude. Also we have not considered the particular form of anisotropy described by a dipole but  a general anisotropy which can be probed by the hemisphere comparison method. 

We have found that for both Pantheon+ and SH0ES SnIa absolute magnitudes the anisotropy level in all bins considered is consistent with the isotropic Monte-Carlo simulated data. However, we have identified a sharp change of the level of anisotropy at the low redshift/distance bins which appears to be rare in the isotropic Monte Carlo simulated data. This change from higher level of anisotropy for absolute magnitudes of SnIa at distances $d<30Mpc$ to lower anisotropy level for absolute magnitudes of SnIa with $d>30Mpc$ is consistent with a scenario assuming an off-center observer in a $20-30Mpc$ bubble with SnIa properties that are distinct from the SnIa properties outside this bubble. 

Such a scenario is also supported by previous studies which have found hints for a transition of the SnIa absolute magnitude and other astrophysical properties at $d\simeq 20Mpc$. This effect could also provide a resolution of the Hubble tension \cite{Perivolaropoulos:2022khd,Perivolaropoulos:2023iqj,Perivolaropoulos:2021bds,Marra:2021fvf,Alestas:2020zol,Alestas:2021nmi} since the value of the SnIa absolute magnitudes is closely connected and degenerate with the measured Hubble parameter $H_0$.

An interesting extension of the present analysis would be to use different cosmological and/or astrophysical probes like Tully-Fisher data to probe possible anisotropies of astrophysics properties in distance bins in the range of $10-50Mpc$. Such studies would provide further tests for the off-center observer interpretation of our results.

{\bf Numerical analysis files:} The Mathematica v13 notebooks and data that lead to the construction of the figures of the paper may be downloaded from \href{https://github.com/leandros11/anisotropy}{this url.}

\section*{Acknowledgements}
This article is based upon work from COST Action CA21136 - Addressing observational tensions in cosmology with systematics and fundamental physics (CosmoVerse), supported by COST (European Cooperation in Science and Technology). This project was also supported by the Hellenic Foundation for Research and Innovation (H.F.R.I.), under the "First call for H.F.R.I. Research Projects to support Faculty members and Researchers and the procurement of high-cost research
equipment Grant" (Project Number: 789).

%\appendix
%\section{Data Used in the Analysis}
%\label{sec:Appendix_A}

\raggedleft
\bibliography{Bibliography}

\end{document}